\documentclass[prl,twocolumn,floatfix]{revtex4-1}
\usepackage{amsfonts}
\usepackage{graphicx}
\usepackage{amsmath}
\usepackage{times}
\usepackage{amssymb}
\usepackage{color}
\usepackage[colorlinks,bookmarks=false,citecolor=blue,linkcolor=red,urlcolor=blue]{hyperref}

\usepackage{soul}

\newcommand{\hamiltonian}{\mathcal{H}}

\newcommand{\sigmax}[1]{\sigma_{#1}^x}

\newcommand{\nop}[1]{n_{#1}}
\newcommand{\ket}[1]{\left| #1 \right\rangle}
\newcommand{\bra}[1]{\left\langle #1 \right|}
\newcommand{\projector}[2]{\ket{#1}\bra{#2}}
\newcommand{\structurefactor}{\mathcal{S}}

\newcommand{\eqeqref}[1]{Eq.~\eqref{#1}}
\newcommand{\figref}[1]{Fig.~\ref{#1}}

\begin{document}

\title{
Floating Phases in One-Dimensional Rydberg Ising Chains
}

\author{Michael Rader}
\affiliation{Institut f\"ur Theoretische Physik, Universit\"at Innsbruck, A-6020 Innsbruck, Austria}
\author{Andreas M. L\"auchli}
\affiliation{Institut f\"ur Theoretische Physik, Universit\"at Innsbruck, A-6020 Innsbruck, Austria}

\date{\today}

\begin{abstract}
We report on the quantitative ground state phase diagram of a van der Waals interacting chain of
Rydberg atoms. These systems are known to host crystalline phases locked to the underlying
lattice as well as a gapped, disordered phase. We locate and characterize a third type of phase, the
so called floating phase, which can be seen as a one-dimensional 'crystalline' phase which is not locked
to the lattice. These phases have been theoretically predicted to exist in the phase diagram, but were 
not reported so far. Our results have been obtained using state-of-the-art numerical tensor network techniques
and pave the way for the experimental exploration of floating phases with existing Rydberg quantum simulators.
\end{abstract}

\maketitle

\paragraph{Introduction ---}
The theoretical and experimental investigation of
strongly interacting quantum matter is one of the central current topics
in condensed matter, quantum optics and even high-energy or nuclear physics.
For this endeavour quantum simulators play an important role, as their controlled
conditions, high tunability and flexibility offer to make powerful inroads towards 
a deeper understanding of quantum matter in and out of equilibrium.

With the recent advent of quantum simulators based on Rydberg atoms 
trapped with optical tweezers or in optical lattices~\cite{Schauss2012,Schauss2015,Labuhn2016,Bernien2017,Zeiher2017,Lienhard2018,Guardado-Sanchez2018,deLeseleuc2018,Keesling2019}
there is a natural interest in systems which can be natively studied using 
these existing platforms. In particular in the recent Harvard experiments~\cite{Bernien2017,Keesling2019,Omran2019}
long one-dimensional chains of Rubidium atoms have been trapped and 
several aspects of quantum many body physics have been studied by 
exciting atoms to interacting Rydberg states. An exemplary study is the 
experimental investigation of quantum Kibble-Zurek dynamics when quenching
the system from a gapped, disordered phase into a gapped, crystalline ordered phase~\cite{Keesling2019}.
This experiment tests the nature of the zero temperature quantum phase transition(s) from the disordered
into the various crystalline phases, as the quantum Kibble-Zurek scaling form
depends on critical exponents, such as the correlation length exponent $\nu$ or the
dynamical critical exponent $z$ of the transitions.

In this context it might come as a surprise, that an accurate quantitative theoretical study of the ground state
phase diagram of the one-dimensional Rydberg model studied in Ref.~\cite{Keesling2019} is currently
lacking, despite a long history of the subject in various contexts~\cite{Fisher1980,Bak1982,Bak1982b,Fendley2004,Weimer2010,Sela2011,Samajdar2018,Whitsitt2018,Chepiga2019,Giudici2019,Verresen2019}. 
It is our goal to close this gap by performing state-of-the-art numerical simulations based on an
infinite-system tensor network approach. We report accurate boundaries of the main crystalline phases and unveil the 
quantitative extent of a so called floating, incommensurate solid phase. 

\begin{figure*}[ht]
	\centering
	\includegraphics[width=0.9\linewidth]{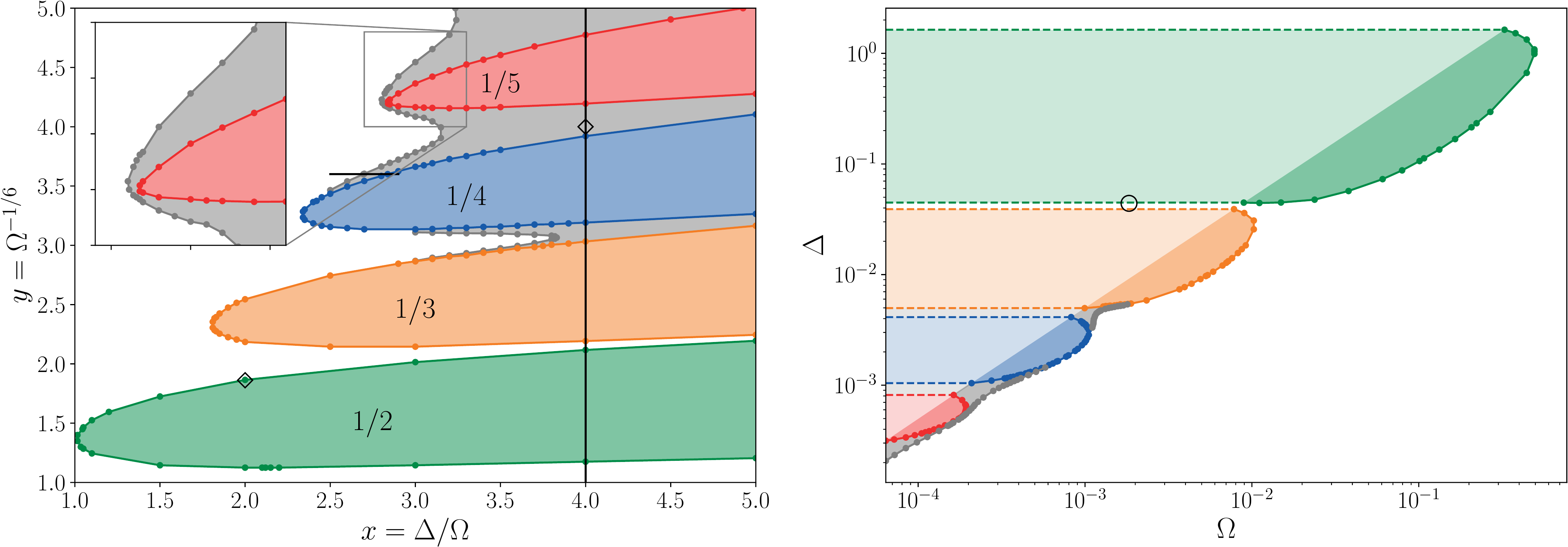}
	\caption{Phase diagram of the model defined in \eqeqref{eq:hamiltonian}. Both panels show the same data, but the left one uses the coordinates $x=\Delta/\Omega$ and $y=\Omega^{-1/6}$, introduced in Ref.~\cite{Keesling2019}. The colored regions indicate $p/q$ crystalline phases with broken translational invariance and the white region corresponds to the gapped, disordered phase. Gray regions indicate the floating phase, which fully coats the $1/5$-lobe.
	Note that the dark parts of the lobes in the right panel are equivalent to the lobes in the left panel and the lighter parts do not contain actual simulation data, but are illustrations of the fact that these
	crystals are expected to extend down to $\Omega=0$.
	Only the dots correspond to actual simulation results: Each of them is the phase transition point obtained from a series of simulations along a horizontal or vertical line. 
	The phase transition lines connecting these dots are guides to the eye. The black lines in the left panel indicate the cuts presented in Fig.~\ref{fig:fouriercuts}, the
	diamond symbols denote the two points shown in Fig.~\ref{fig:eexiscaling}, and the circle in the right panel indicates the location of the cut presented in Fig.~\ref{fig:fouriercuts2}.}
	\label{fig:phasediagram}
\end{figure*}

\begin{figure}[b]
	\centering
	\includegraphics[width=0.85\linewidth]{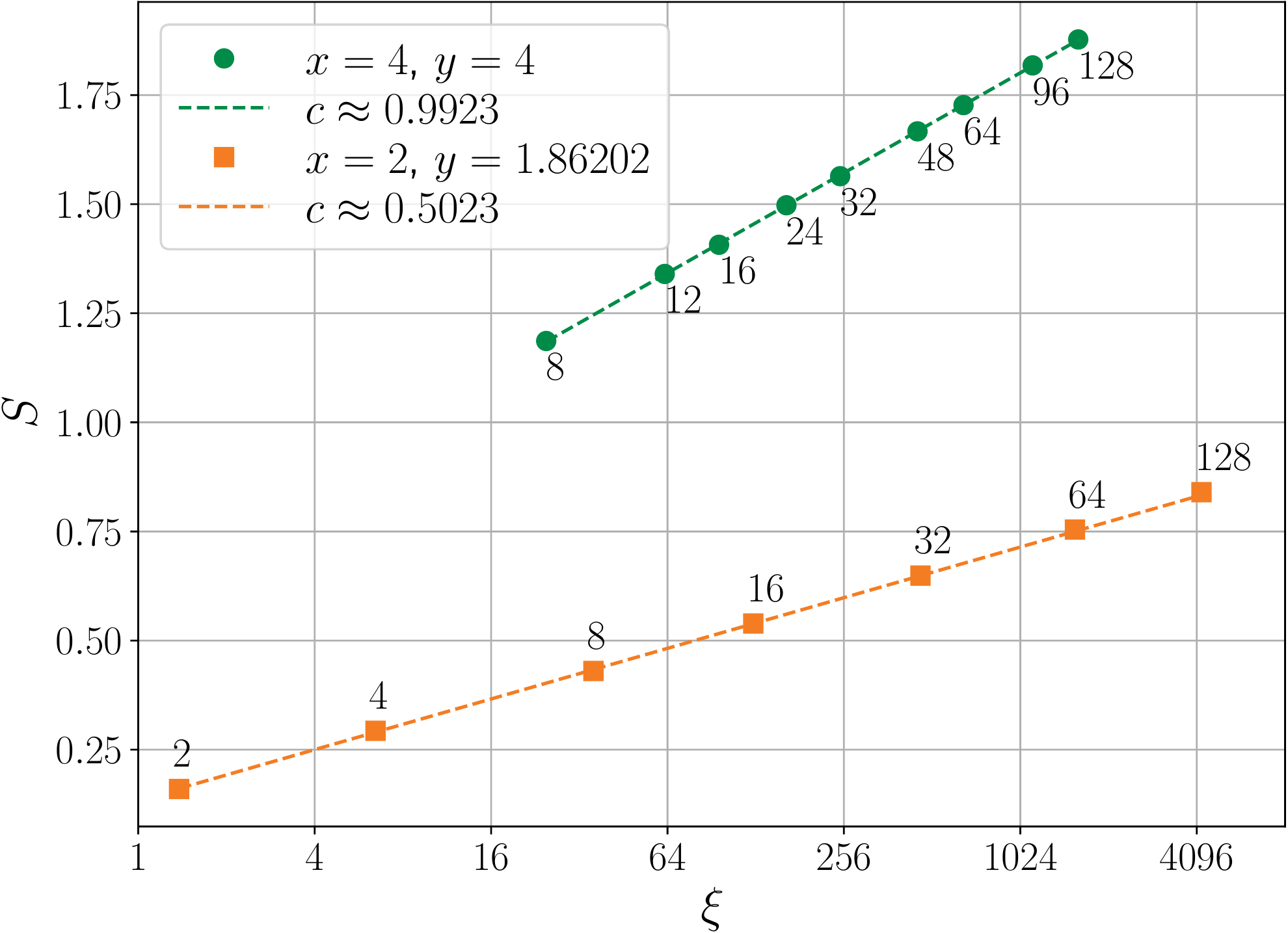}
	\caption{Scaling of the entanglement entropy $S$ as a function of the correlation length $\xi$ for a system in the floating phase at $x=4$, $y=4$ (green circles) and for a system close the Ising phase transition between the $1/2$-lobe and the disordered phase at $x=2$, $y=1.86202$ (orange squares). Labels next to the points indicate the used bond dimensions $\chi$. The dashed lines show fits according to the model $S = \frac{c}{6} \log(\xi)+\mathrm{const.}$, from which estimates for the central charge $c$ are extracted.}
	\label{fig:eexiscaling}
\end{figure}

\paragraph{Model and expected phases ---}
We study an infinite chain of Rydberg atoms described by the Hamiltonian,
\begin{equation}
	\label{eq:hamiltonian}
	\hamiltonian{} = 
		- \frac{\Omega}{2} \sum\limits_{j} \sigmax{j}
		- \Delta \sum\limits_{j} \nop{j}
		+ V \sum\limits_{j<l} \frac{\nop{j} \nop{l}}{(l-j)^6}
	\text{.}
\end{equation}
The local Hilbert space for an atom on site $j$ is spanned by the ground state $\ket{g_j}$ and the excited Rydberg state $\ket{r_j}$.
The single-site operators are $\sigmax{j} = \projector{g_j}{r_j} + \projector{r_j}{g_j}$ and $\nop{j} = \projector{r_j}{r_j}$. 
The parameters $\Omega$ and $\Delta$ are called Rabi frequency and laser detuning.
Throughout this work we use a unit lattice spacing and $V=1$, such that the corresponding Rydberg-Rydberg interaction gives an energy penalty for excited Rydberg atoms, which decreases with the distance $r$ between these atoms with a characteristic van der Waals $1/r^6$ power law. 
This Rydberg chain can be mapped onto a spin chain model~\footnote{This interacting Rydberg chain is equivalent to a spin one-half chain with power-law antiferromagnetic Ising interactions, a transverse field $\frac{\Omega}{2}$, and a longitudinal field $\Delta - \zeta(6)$. This mapping is useful to understand that the phase diagram features a symmetry, $(\Delta, n) \mapsto (2 \zeta(6) - \Delta, 1 - n)$,
where $\zeta(n)$ is the Riemann Zeta function.}, and
it is therefore sufficient to explore the physics for $\Delta \leq \zeta(6)$, as the remaining region can be obtained by virtue of the mapping.

For $\Omega = 0$, i.e.~in the absence of quantum fluctuations, the model is purely classical and the ground states form a complete Devil's staircase of crystalline phases,
as shown by Bak and Bruinsma~\cite{Bak1982}. These translation-symmetry breaking crystals are characterized by coprime integers $p$ and $q$, where $q$ denotes the size of the unit cell and $p$ the number of atoms in the Rydberg state within the unit cell. The atoms in the Rydberg state are located as far apart as possible, given $p$ and $q$. The extent in detuning of a crystal 
only depends on $q$~\cite{Burnell2009}, so that e.g.~a crystal with $p/q=1/5$ has the same width as the one with $p/q=2/5$ in the limit $\Omega=0$. Note that due to the rapid $1/r^6$
decay of the interactions, the widths of the plateaux shrink rapidly with increasing $q$.

For large detuning and/or large Rabi frequency the interaction can be neglected and the ground state corresponds in essence to a product state of the single site ground state
dictated by the local terms. This ground state breaks no symmetry, has a large excitation gap and rapidly decaying correlations. 

In between the two limits considered there are only a few quantitative results so far for the van der Waals interacting case~\cite{Weimer2010,Sela2011,Keesling2019} (see 
Refs~\cite{Deng2005,Schachenmayer2010,Nebendahl2015} for results for the dipolar $1/r^3$ case).
One generally expects the crystalline phases to be stable upon the addition of the Rabi drive $\Omega$, up to a strength which is in line with its width in detuning. The plethora of proposed scenarios of the quantum fluctuation induced melting of the crystalline phases is however quite diverse~\cite{Fendley2004,Sela2011,Whitsitt2018,Chepiga2019}, and requires an accurate numerical investigation. 

In Ref.~\cite{Weimer2010} it was advocated in the Rydberg context, that in general the transition from a commensurate $p/q$ crystal to the gapped, disordered phase proceeds 
via an extended, incommensurate and gapless phase, the so called {\em floating phase}. Such phases have a long history~\cite{Pokrovsky1979,Bak1982b} and have been introduced
in the context of commensurate-incommensurate transitions in solids. In modern parlance these phases are Luttinger liquids~\cite{Haldane1981,GiamarchiBook}, whose characteristics, such as the 
Luttinger parameter $K$ and the "Fermi" momentum $k_F$ can depend on microscopic parameters, here $\Omega$ and $\Delta$. Other aspects however, such as the central charge $c=1$
and the qualitative power-law decay of correlations are universal. In our work we specifically search for these predicted phases, because they have not been reported in numerical simulations so far,
other than in tiny slivers around the $p/q=1/3$ crystal in a related, albeit different, model~\cite{Fendley2004,Chepiga2019,Giudici2019}. We find that floating phases feature rather prominently
in the Rydberg chain phase diagram, even for experimentally accessible parameter ranges, see Fig.~\ref{fig:phasediagram}.

\begin{figure*}[ht]
	\centering
	\includegraphics[width=0.45\linewidth]{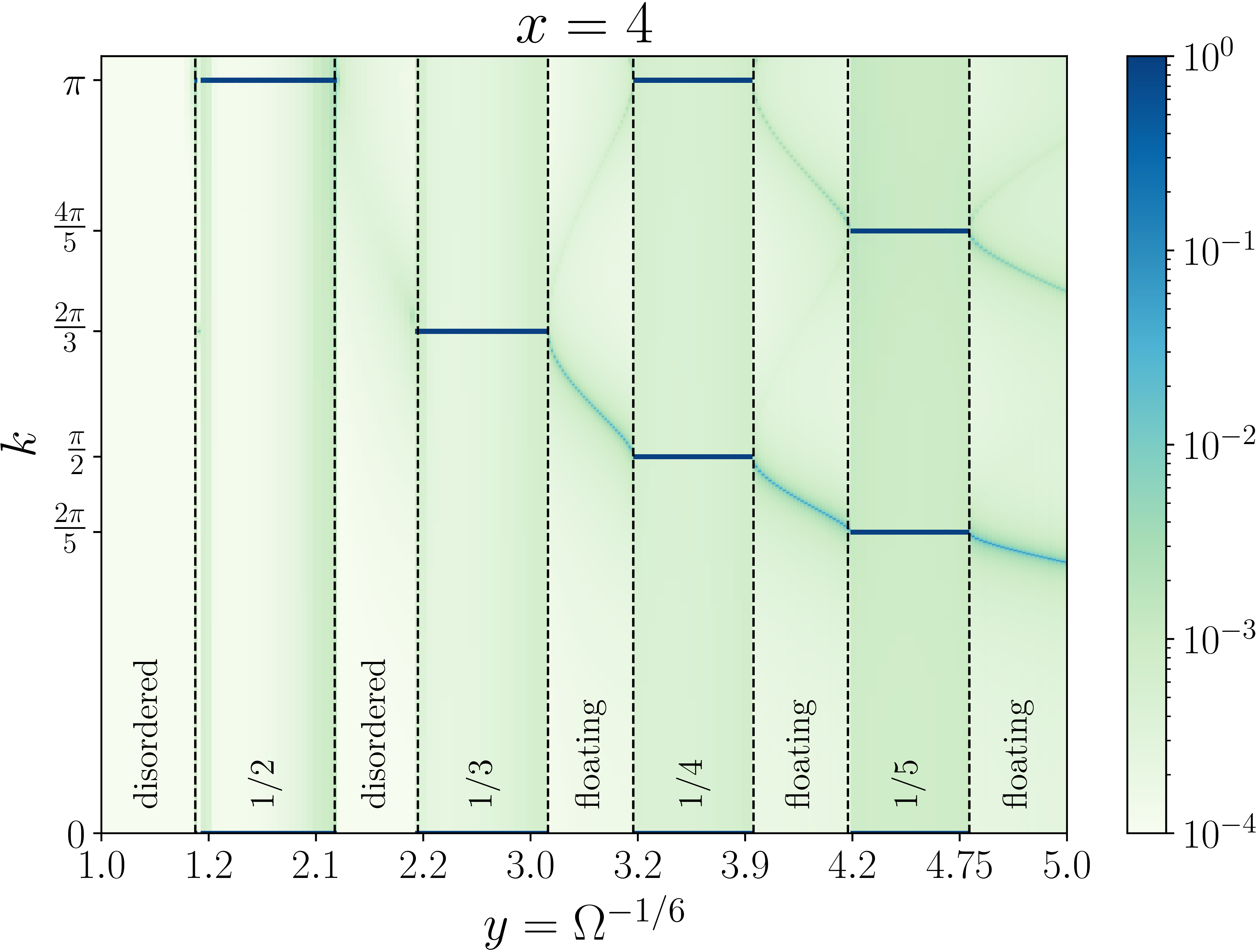} 
	\hspace{10mm} 
	\includegraphics[width=0.45\linewidth]{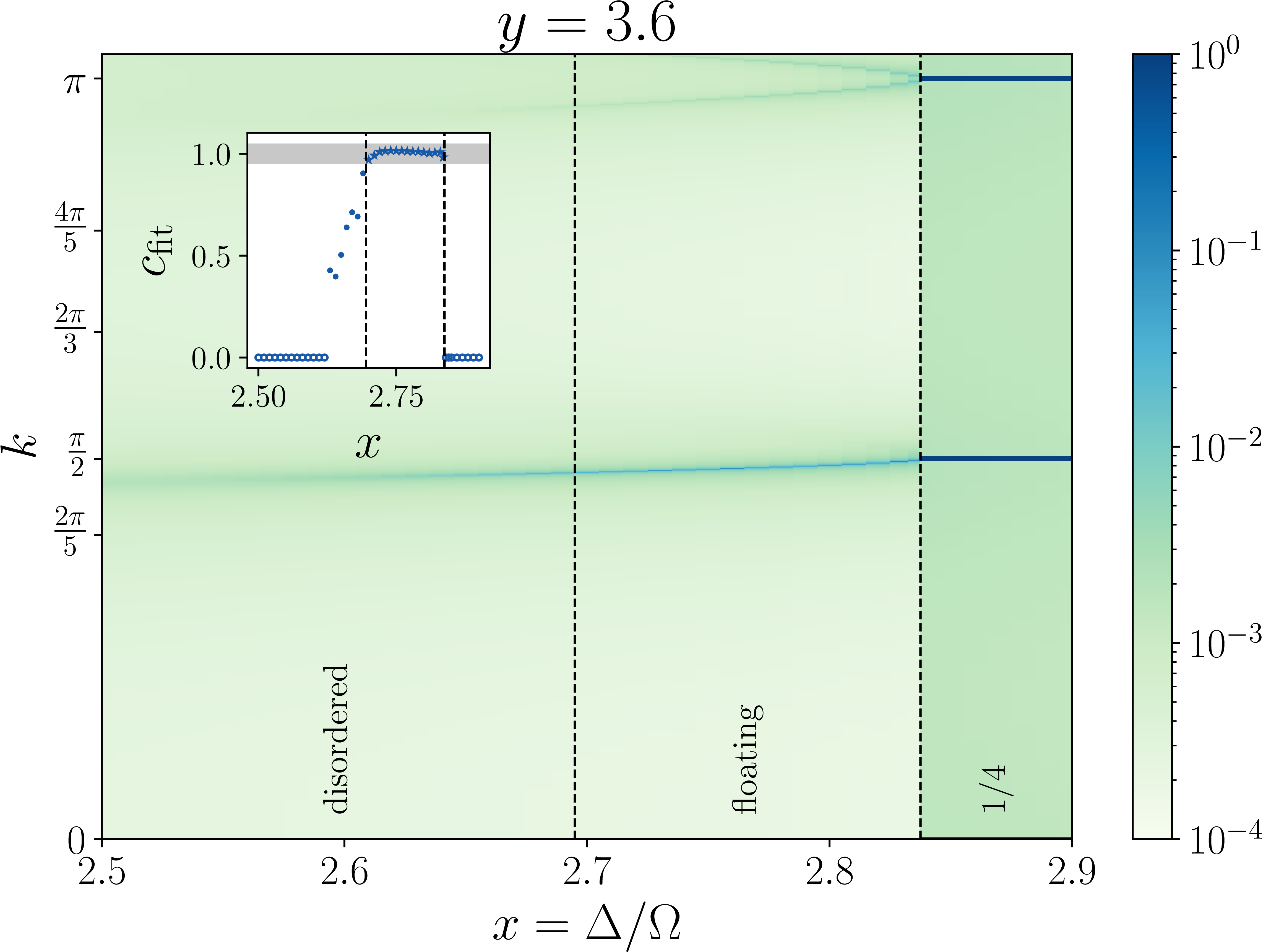}
	\caption{Vertical strips show the structure factor $\structurefactor(k)$ as defined in \eqeqref{eq:structurefactor} for cuts through the phase diagram at $x=4$ ($y=3.6$) in the left (right) panel. 
	As the structure factors are symmetric around the Brillouin zone boundary at $k=\pi$, only one half of the zone is shown.
	In an ordered phase $p / q$ one observes Bragg peaks at $k=2\pi n /q$, $n=0,\ldots,q-1$. 	
	It should be noted that the finite width of these peaks in the plot are for improved readability. 
	The floating phases show power-law diverging peaks at drifting wave vectors $k$, and weaker peaks at higher harmonics of the wave vector $k_n=n \, k$, $n\in\mathbb{Z}$, 
	which interpolate between the peaks of the crystalline phases. In the disordered phases 
	the structure factor is not divergent at finite wave vector and has broad peaks. In the inset of the right panel one can observe the extended central charge $c=1$ region, which characterizes the floating phase.} 
	\label{fig:fouriercuts}
\end{figure*}

\paragraph{Method ---} In this work we use infinite matrix product states (iMPS) \cite{Vidal2007} for studying the ground state of our model.
An iMPS is a variational tensor network ansatz for an infinitely long chain with a periodically repeated $L_\mathrm{uc}$-site unit cell and a bond dimension $\chi$, which systematically limits the entanglement entropy $S$ to 
an area law, i.e. $S \leq \log(\chi)$.
We optimize the energy of the iMPS using the infinite density matrix renormalization group (iDMRG) algorithm~\cite{McCulloch2008,Kjall2013} as provided by the TeNPy library~\cite{tenpy}.
iDMRG requires the Hamiltonian to be written in the form of a so-called infinite matrix product operator (iMPO). Unfortunately, one cannot directly formulate such an iMPO for power-law long-range interactions, but accurate
decompositions into a sum of exponentials are a known way to proceed~\cite{Crosswhite2008,Pirvu2010,Nebendahl2013}. Here we use a decomposition comprising a sum of ten exponentials. The choice of $L_\mathrm{uc}$ is crucial for iDMRG:  If an incompatible unit cell is chosen, 
i.e. $L_\mathrm{uc}$ is not a multiple of $q$ for a system where the true ground state is a $p/q$ crystal,
iDMRG will be numerically unstable or yield an iMPS in a different close-by physical phase.
Therefore, this effect can be used to stabilize the two phases close to a phase transition with different choices of $L_\mathrm{uc}$ 
and finally find the true ground state depending on the obtained energies for $\chi \rightarrow \infty$. For gapped phases the energy can easily be converged in 
the iMPS bond dimension $\chi$, but for gapless phases we have to compute the correlation length $\xi$ and extrapolate the 
energy as a function of $1/\xi^2$, which is the expected Casimir scaling. Bond dimensions $\chi$ ranging from $4$ to $256$ are 
sufficient to obtain the energies reliably~\footnote{We observed that 
particularly in simulations of crystalline phases, the smallest eigenvalue of the effective iDMRG-Hamiltonian 
does not coincide with the true energy of the state, although all other observables converge properly. Therefore, as a cross-check, 
we evaluate the energy with interactions truncated to a few hundred sites.}.

The $p/q$ crystalline phases can be discriminated from the homogeneous, disordered phase by detecting a density imbalance within the unit cell (here $\delta n \gtrsim 0.05$), 
as the iDMRG algorithm generically converges to a minimally entangled instance within the $q$-fold degenerate ground state manifold,
and Schr\"odinger cat-like superpositions are avoided. 

If a system is homogeneous, we still have to discriminate between disordered and floating phases. For the floating, gapless phase or isolated critical points, 
we can relate the correlation length $\xi$ and entanglement entropy $S$, 
\begin{equation}
	\label{eq:eexisclaing}
	S  = \frac{c}{6} \log(\xi) + \mathrm{const.}
	\text{,}
\end{equation}
and extract the central charge $c$ \cite{Pollmann2009} as shown for two exemplary cases in \figref{fig:eexiscaling}. For automatic classification we use the criterion 
$c = 1 \pm 5 \%$ to declare the system as floating. Due to the Kosterlitz-Thouless nature of the quantum phase transition from the disordered to the floating phase 
and the concomitant logarithmic corrections, the extent of the floating phase might be slightly overestimated using this criterion.

We further define the structure factor,
\begin{equation}
	\label{eq:structurefactor}
	\structurefactor(k) \propto 
	\left|
		\sum\limits_{r=0}^{R}
		\langle n_0 n_{0+r} \rangle
		e^{
			-i k r 
		}
	\right|
	\text{,}
\end{equation}
where $R$ denotes the number of sites included in the Fourier transform~\footnote{Here we use $R=300$ in the crystalline phases and $R=10000$ for the remaining.
phases.}

\begin{figure}[b]
\centering
\includegraphics[width=0.85\linewidth]{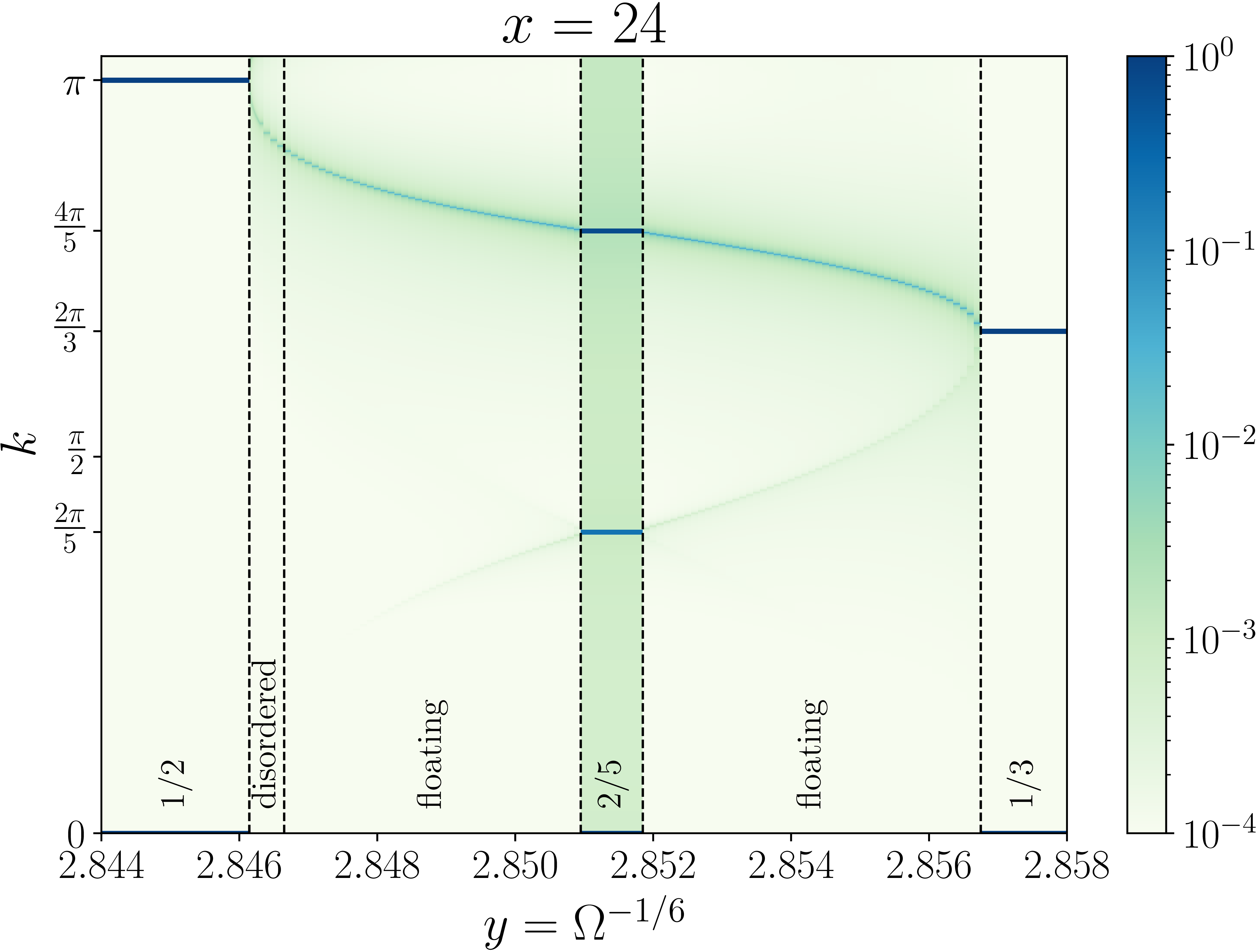}
\caption{Vertical strips show the structure factor $\structurefactor(k)$ as defined in \eqeqref{eq:structurefactor} for a cut through the phase diagram at $x=24$. In addition to the characteristic features already discussed for \figref{fig:fouriercuts}, for this value of $x$ also the $2/5$ plateau appears.}
\label{fig:fouriercuts2}
\end{figure}

\paragraph{Phase diagram with floating phases ---}
In \figref{fig:phasediagram} we show the phase diagram obtained with our simulations. The left panel shows the data using the coordinates, $x = \Delta/\Omega$ and $y = \Omega^{-1/6}$, as introduced in Ref.~\cite{Keesling2019}. The phase transition points are obtained by running a series of simulations for either constant $x$ or $y$ and the phase transition lines connecting these points are guides to the eye.
In this first plot we have restricted ourselves to the region $1\leq x,y\leq 5$. There we find ordered phases with fractional fillings $1/q$, with $q=2,3,4,5$ as indicated by the colored lobes. 
We further find sizable floating phases indicated by the gray regions between the ordered phases, which have not been reported previously~\cite{Keesling2019}. 
For larger values of $y$ the floating phase clearly approaches the tips of the ordered lobes and the $1/5$-lobe appears fully coated by the floating phase. Based on this observation and the arguments 
in Ref.~\cite{Weimer2010,Sela2011} we believe that all $1/q$-lobes for $q \geq 5$ are fully immersed into the floating phase. Note that there is a floating phase in between the $1/3$ and the $1/4$ lobe, which
appears at experimentally accessible values of $\Omega$ and $\Delta$~\cite{Keesling2019}, so they could be studied using the existing experiments.

The two black lines in the left panel of \figref{fig:phasediagram} indicate two cuts through the phase diagram at fixed $x=4$ and fixed $y=3.6$, respectively, for which structure factors $\structurefactor(k)$ 
are presented in \figref{fig:fouriercuts}. Crystalline phases $p/q$ exhibit infinitely sharp Bragg (i.e.~$\delta$-function) peaks at wave vector $k=2\pi n /q$, $n=0,\ldots,q-1$. 
In the floating phases the real-space Rydberg density correlations decay as a power-law $1/r^\alpha$, with an exponent $\alpha\leq 1/4$~\cite{Weimer2010,Sela2011}. 
These correlations translate into power-law diverging peaks in the structure factor. Furthermore the wavevector of the peaks of the structure factor are not locked to commensurate values in the floating phase, 
and they indeed show a pronounced dependence on the parameters in the phase diagram. In the left panel of \figref{fig:fouriercuts} one can see how the floating phase interpolates between and touches the $1/3$, $1/4$ and $1/5$ crystals. One expects the wave vector corresponding to the peaks in the structure factor to approach the commensurate values with a square-root singularity~\cite{Pokrovsky1979}. For the $1/3$ crystal this is nicely visible (see also \figref{fig:fouriercuts2} below.). For higher $q$ this behavior seems restricted to a more narrow window in the vicinity of the crystal, similar to observations in frustrated spin ladders~\cite{Fouet2006}.

 On the scale presented we do not observe a floating phase to the small $y$ side of the $1/3$ crystal, although according to recent results in a related model, a very thin sliver is likely to occur~\cite{Chepiga2019}. For larger values of $x$ presented below, we will also observe a sizeable floating phase on that side of the $1/3$ crystal. Finally note that we have not observed a floating phase touching the $1/2$ lobe, as predicted in Ref.~\cite{Sela2011}. Indeed in \figref{fig:eexiscaling} we clearly observe the central charge $c=1/2$ of the Ising universality class expected for direct transition from the disordered phase into the $1/2$ crystal. For the perpendicular cut at $y=3.6$ (right panel of \figref{fig:fouriercuts}) one can observe the successive sharpening of the primary peak as one moves from the gapped, disordered phase into the floating phase and finally 
into the $1/4$ crystal phase. In the inset one can observe the extended central charge $c=1$ region, which characterizes the floating phase. 

So far, we have only observed crystalline phases of type $1/q$ in the phase diagram, however in the same range of $y$, but for larger values of $x$ one can also find other plateaux of the complete devil's staircase at $\Omega=0$. To illustrate this, we show the structure factor for a cut at $x=24$ in \figref{fig:fouriercuts2}, which is beyond the scope of the phase diagram in the left panel of \figref{fig:phasediagram}. For this value of $x$ the $2/5$ plateau appears between the plateaux $1/2$ and $1/3$, which is coated by a sizeable floating phase, which however stops shortly before touching the $1/2$ crystal lobe.

\paragraph{Conclusion ---} We have established the quantitative ground state phase diagram 
of a van der Waals interacting chain of Rydberg atoms. Our main result is to locate and characterize 
theso called floating phases, which can be seen as one-dimensional 'crystalline' phases which are not locked
to the lattice. Our results pave the way for the experimental exploration of floating phases with existing 
Rydberg quantum simulators.

\begin{acknowledgments}
We acknowledge helpful discussions with M.~Dalmonte, M.~Garst, J.~Hauschild,
F.~Pollmann, and R.~Verresen. AML acknowledges V. Nebendahl for collaborating on 
previous work on the dipolar case, which can be found in Ref.~\cite{Nebendahl2015}.
We  acknowledge  support by the Austrian Science Fund FWF within 
the DK-ALM (W1259-N27) and the SFB FoQus (F-4018). 
The computational results presented have been achieved in part using the HPC 
infrastructure LEO of the University of Innsbruck. The computational results presented 
have been achieved in part using the Vienna Scientific Cluster (VSC).
\end{acknowledgments}


\begin{thebibliography}{40}%
\makeatletter
\providecommand \@ifxundefined [1]{%
 \@ifx{#1\undefined}
}%
\providecommand \@ifnum [1]{%
 \ifnum #1\expandafter \@firstoftwo
 \else \expandafter \@secondoftwo
 \fi
}%
\providecommand \@ifx [1]{%
 \ifx #1\expandafter \@firstoftwo
 \else \expandafter \@secondoftwo
 \fi
}%
\providecommand \natexlab [1]{#1}%
\providecommand \enquote  [1]{``#1''}%
\providecommand \bibnamefont  [1]{#1}%
\providecommand \bibfnamefont [1]{#1}%
\providecommand \citenamefont [1]{#1}%
\providecommand \href@noop [0]{\@secondoftwo}%
\providecommand \href [0]{\begingroup \@sanitize@url \@href}%
\providecommand \@href[1]{\@@startlink{#1}\@@href}%
\providecommand \@@href[1]{\endgroup#1\@@endlink}%
\providecommand \@sanitize@url [0]{\catcode `\\12\catcode `\$12\catcode
  `\&12\catcode `\#12\catcode `\^12\catcode `\_12\catcode `\%12\relax}%
\providecommand \@@startlink[1]{}%
\providecommand \@@endlink[0]{}%
\providecommand \url  [0]{\begingroup\@sanitize@url \@url }%
\providecommand \@url [1]{\endgroup\@href {#1}{\urlprefix }}%
\providecommand \urlprefix  [0]{URL }%
\providecommand \Eprint [0]{\href }%
\providecommand \doibase [0]{https://doi.org/}%
\providecommand \selectlanguage [0]{\@gobble}%
\providecommand \bibinfo  [0]{\@secondoftwo}%
\providecommand \bibfield  [0]{\@secondoftwo}%
\providecommand \translation [1]{[#1]}%
\providecommand \BibitemOpen [0]{}%
\providecommand \bibitemStop [0]{}%
\providecommand \bibitemNoStop [0]{.\EOS\space}%
\providecommand \EOS [0]{\spacefactor3000\relax}%
\providecommand \BibitemShut  [1]{\csname bibitem#1\endcsname}%
\let\auto@bib@innerbib\@empty
\bibitem [{\citenamefont {Schau{\ss}}\ \emph {et~al.}(2012)\citenamefont
  {Schau{\ss}}, \citenamefont {Cheneau}, \citenamefont {Endres}, \citenamefont
  {Fukuhara}, \citenamefont {Hild}, \citenamefont {Omran}, \citenamefont
  {Pohl}, \citenamefont {Gross}, \citenamefont {Kuhr},\ and\ \citenamefont
  {Bloch}}]{Schauss2012}%
  \BibitemOpen
  \bibfield  {author} {\bibinfo {author} {\bibfnamefont {P.}~\bibnamefont
  {Schau{\ss}}}, \bibinfo {author} {\bibfnamefont {M.}~\bibnamefont {Cheneau}},
  \bibinfo {author} {\bibfnamefont {M.}~\bibnamefont {Endres}}, \bibinfo
  {author} {\bibfnamefont {T.}~\bibnamefont {Fukuhara}}, \bibinfo {author}
  {\bibfnamefont {S.}~\bibnamefont {Hild}}, \bibinfo {author} {\bibfnamefont
  {A.}~\bibnamefont {Omran}}, \bibinfo {author} {\bibfnamefont
  {T.}~\bibnamefont {Pohl}}, \bibinfo {author} {\bibfnamefont {C.}~\bibnamefont
  {Gross}}, \bibinfo {author} {\bibfnamefont {S.}~\bibnamefont {Kuhr}},\ and\
  \bibinfo {author} {\bibfnamefont {I.}~\bibnamefont {Bloch}},\ }\bibfield
  {title} {\bibinfo {title} {{Observation of spatially ordered structures in a
  two-dimensional Rydberg gas}},\ }\href {https://doi.org/10.1038/nature11596}
  {\bibfield  {journal} {\bibinfo  {journal} {Nature}\ }\textbf {\bibinfo
  {volume} {491}},\ \bibinfo {pages} {87} (\bibinfo {year} {2012})}\BibitemShut
  {NoStop}%
\bibitem [{\citenamefont {Schau{\ss}}\ \emph {et~al.}(2015)\citenamefont
  {Schau{\ss}}, \citenamefont {Zeiher}, \citenamefont {Fukuhara}, \citenamefont
  {Hild}, \citenamefont {Cheneau}, \citenamefont {Macr{\`\i}}, \citenamefont
  {Pohl}, \citenamefont {Bloch},\ and\ \citenamefont {Gross}}]{Schauss2015}%
  \BibitemOpen
  \bibfield  {author} {\bibinfo {author} {\bibfnamefont {P.}~\bibnamefont
  {Schau{\ss}}}, \bibinfo {author} {\bibfnamefont {J.}~\bibnamefont {Zeiher}},
  \bibinfo {author} {\bibfnamefont {T.}~\bibnamefont {Fukuhara}}, \bibinfo
  {author} {\bibfnamefont {S.}~\bibnamefont {Hild}}, \bibinfo {author}
  {\bibfnamefont {M.}~\bibnamefont {Cheneau}}, \bibinfo {author} {\bibfnamefont
  {T.}~\bibnamefont {Macr{\`\i}}}, \bibinfo {author} {\bibfnamefont
  {T.}~\bibnamefont {Pohl}}, \bibinfo {author} {\bibfnamefont {I.}~\bibnamefont
  {Bloch}},\ and\ \bibinfo {author} {\bibfnamefont {C.}~\bibnamefont {Gross}},\
  }\bibfield  {title} {\bibinfo {title} {Crystallization in ising quantum
  magnets},\ }\href {https://doi.org/10.1126/science.1258351} {\bibfield
  {journal} {\bibinfo  {journal} {Science}\ }\textbf {\bibinfo {volume}
  {347}},\ \bibinfo {pages} {1455} (\bibinfo {year} {2015})}\BibitemShut
  {NoStop}%
\bibitem [{\citenamefont {Labuhn}\ \emph {et~al.}(2016)\citenamefont {Labuhn},
  \citenamefont {Barredo}, \citenamefont {Ravets}, \citenamefont
  {de~L{\'e}s{\'e}leuc}, \citenamefont {Macr{\`\i}}, \citenamefont {Lahaye},\
  and\ \citenamefont {Browaeys}}]{Labuhn2016}%
  \BibitemOpen
  \bibfield  {author} {\bibinfo {author} {\bibfnamefont {H.}~\bibnamefont
  {Labuhn}}, \bibinfo {author} {\bibfnamefont {D.}~\bibnamefont {Barredo}},
  \bibinfo {author} {\bibfnamefont {S.}~\bibnamefont {Ravets}}, \bibinfo
  {author} {\bibfnamefont {S.}~\bibnamefont {de~L{\'e}s{\'e}leuc}}, \bibinfo
  {author} {\bibfnamefont {T.}~\bibnamefont {Macr{\`\i}}}, \bibinfo {author}
  {\bibfnamefont {T.}~\bibnamefont {Lahaye}},\ and\ \bibinfo {author}
  {\bibfnamefont {A.}~\bibnamefont {Browaeys}},\ }\bibfield  {title} {\bibinfo
  {title} {{Tunable two-dimensional arrays of single Rydberg atoms for
  realizing quantum Ising models}},\ }\href
  {https://doi.org/10.1038/nature18274} {\bibfield  {journal} {\bibinfo
  {journal} {Nature}\ }\textbf {\bibinfo {volume} {534}},\ \bibinfo {pages}
  {667} (\bibinfo {year} {2016})}\BibitemShut {NoStop}%
\bibitem [{\citenamefont {Bernien}\ \emph {et~al.}(2017)\citenamefont
  {Bernien}, \citenamefont {Schwartz}, \citenamefont {Keesling}, \citenamefont
  {Levine}, \citenamefont {Omran}, \citenamefont {Pichler}, \citenamefont
  {Choi}, \citenamefont {Zibrov}, \citenamefont {Endres}, \citenamefont
  {Greiner}, \citenamefont {Vuleti{\'c}},\ and\ \citenamefont
  {Lukin}}]{Bernien2017}%
  \BibitemOpen
  \bibfield  {author} {\bibinfo {author} {\bibfnamefont {H.}~\bibnamefont
  {Bernien}}, \bibinfo {author} {\bibfnamefont {S.}~\bibnamefont {Schwartz}},
  \bibinfo {author} {\bibfnamefont {A.}~\bibnamefont {Keesling}}, \bibinfo
  {author} {\bibfnamefont {H.}~\bibnamefont {Levine}}, \bibinfo {author}
  {\bibfnamefont {A.}~\bibnamefont {Omran}}, \bibinfo {author} {\bibfnamefont
  {H.}~\bibnamefont {Pichler}}, \bibinfo {author} {\bibfnamefont
  {S.}~\bibnamefont {Choi}}, \bibinfo {author} {\bibfnamefont {A.~S.}\
  \bibnamefont {Zibrov}}, \bibinfo {author} {\bibfnamefont {M.}~\bibnamefont
  {Endres}}, \bibinfo {author} {\bibfnamefont {M.}~\bibnamefont {Greiner}},
  \bibinfo {author} {\bibfnamefont {V.}~\bibnamefont {Vuleti{\'c}}},\ and\
  \bibinfo {author} {\bibfnamefont {M.~D.}\ \bibnamefont {Lukin}},\ }\bibfield
  {title} {\bibinfo {title} {Probing many-body dynamics on a 51-atom quantum
  simulator},\ }\href {https://doi.org/10.1038/nature24622} {\bibfield
  {journal} {\bibinfo  {journal} {Nature}\ }\textbf {\bibinfo {volume} {551}},\
  \bibinfo {pages} {579} (\bibinfo {year} {2017})}\BibitemShut {NoStop}%
\bibitem [{\citenamefont {Zeiher}\ \emph {et~al.}(2017)\citenamefont {Zeiher},
  \citenamefont {Choi}, \citenamefont {Rubio-Abadal}, \citenamefont {Pohl},
  \citenamefont {van Bijnen}, \citenamefont {Bloch},\ and\ \citenamefont
  {Gross}}]{Zeiher2017}%
  \BibitemOpen
  \bibfield  {author} {\bibinfo {author} {\bibfnamefont {J.}~\bibnamefont
  {Zeiher}}, \bibinfo {author} {\bibfnamefont {J.-y.}\ \bibnamefont {Choi}},
  \bibinfo {author} {\bibfnamefont {A.}~\bibnamefont {Rubio-Abadal}}, \bibinfo
  {author} {\bibfnamefont {T.}~\bibnamefont {Pohl}}, \bibinfo {author}
  {\bibfnamefont {R.}~\bibnamefont {van Bijnen}}, \bibinfo {author}
  {\bibfnamefont {I.}~\bibnamefont {Bloch}},\ and\ \bibinfo {author}
  {\bibfnamefont {C.}~\bibnamefont {Gross}},\ }\bibfield  {title} {\bibinfo
  {title} {{Coherent Many-Body Spin Dynamics in a Long-Range Interacting Ising
  Chain}},\ }\href {https://doi.org/10.1103/PhysRevX.7.041063} {\bibfield
  {journal} {\bibinfo  {journal} {Phys. Rev. X}\ }\textbf {\bibinfo {volume}
  {7}},\ \bibinfo {pages} {041063} (\bibinfo {year} {2017})}\BibitemShut
  {NoStop}%
\bibitem [{\citenamefont {Lienhard}\ \emph {et~al.}(2018)\citenamefont
  {Lienhard}, \citenamefont {de~L\'es\'eleuc}, \citenamefont {Barredo},
  \citenamefont {Lahaye}, \citenamefont {Browaeys}, \citenamefont {Schuler},
  \citenamefont {Henry},\ and\ \citenamefont {L\"auchli}}]{Lienhard2018}%
  \BibitemOpen
  \bibfield  {author} {\bibinfo {author} {\bibfnamefont {V.}~\bibnamefont
  {Lienhard}}, \bibinfo {author} {\bibfnamefont {S.}~\bibnamefont
  {de~L\'es\'eleuc}}, \bibinfo {author} {\bibfnamefont {D.}~\bibnamefont
  {Barredo}}, \bibinfo {author} {\bibfnamefont {T.}~\bibnamefont {Lahaye}},
  \bibinfo {author} {\bibfnamefont {A.}~\bibnamefont {Browaeys}}, \bibinfo
  {author} {\bibfnamefont {M.}~\bibnamefont {Schuler}}, \bibinfo {author}
  {\bibfnamefont {L.-P.}\ \bibnamefont {Henry}},\ and\ \bibinfo {author}
  {\bibfnamefont {A.~M.}\ \bibnamefont {L\"auchli}},\ }\bibfield  {title}
  {\bibinfo {title} {{Observing the Space- and Time-Dependent Growth of
  Correlations in Dynamically Tuned Synthetic Ising Models with
  Antiferromagnetic Interactions}},\ }\href
  {https://doi.org/10.1103/PhysRevX.8.021070} {\bibfield  {journal} {\bibinfo
  {journal} {Phys. Rev. X}\ }\textbf {\bibinfo {volume} {8}},\ \bibinfo {pages}
  {021070} (\bibinfo {year} {2018})}\BibitemShut {NoStop}%
\bibitem [{\citenamefont {Guardado-Sanchez}\ \emph {et~al.}(2018)\citenamefont
  {Guardado-Sanchez}, \citenamefont {Brown}, \citenamefont {Mitra},
  \citenamefont {Devakul}, \citenamefont {Huse}, \citenamefont {Schau\ss{}},\
  and\ \citenamefont {Bakr}}]{Guardado-Sanchez2018}%
  \BibitemOpen
  \bibfield  {author} {\bibinfo {author} {\bibfnamefont {E.}~\bibnamefont
  {Guardado-Sanchez}}, \bibinfo {author} {\bibfnamefont {P.~T.}\ \bibnamefont
  {Brown}}, \bibinfo {author} {\bibfnamefont {D.}~\bibnamefont {Mitra}},
  \bibinfo {author} {\bibfnamefont {T.}~\bibnamefont {Devakul}}, \bibinfo
  {author} {\bibfnamefont {D.~A.}\ \bibnamefont {Huse}}, \bibinfo {author}
  {\bibfnamefont {P.}~\bibnamefont {Schau\ss{}}},\ and\ \bibinfo {author}
  {\bibfnamefont {W.~S.}\ \bibnamefont {Bakr}},\ }\bibfield  {title} {\bibinfo
  {title} {{Probing the Quench Dynamics of Antiferromagnetic Correlations in a
  2D Quantum Ising Spin System}},\ }\href
  {https://doi.org/10.1103/PhysRevX.8.021069} {\bibfield  {journal} {\bibinfo
  {journal} {Phys. Rev. X}\ }\textbf {\bibinfo {volume} {8}},\ \bibinfo {pages}
  {021069} (\bibinfo {year} {2018})}\BibitemShut {NoStop}%
\bibitem [{\citenamefont {{de L{\'e}s{\'e}leuc}}\ \emph
  {et~al.}(2018)\citenamefont {{de L{\'e}s{\'e}leuc}}, \citenamefont
  {{Lienhard}}, \citenamefont {{Scholl}}, \citenamefont {{Barredo}},
  \citenamefont {{Weber}}, \citenamefont {{Lang}}, \citenamefont
  {{B{\"u}chler}}, \citenamefont {{Lahaye}},\ and\ \citenamefont
  {{Browaeys}}}]{deLeseleuc2018}%
  \BibitemOpen
  \bibfield  {author} {\bibinfo {author} {\bibfnamefont {S.}~\bibnamefont {{de
  L{\'e}s{\'e}leuc}}}, \bibinfo {author} {\bibfnamefont {V.}~\bibnamefont
  {{Lienhard}}}, \bibinfo {author} {\bibfnamefont {P.}~\bibnamefont
  {{Scholl}}}, \bibinfo {author} {\bibfnamefont {D.}~\bibnamefont {{Barredo}}},
  \bibinfo {author} {\bibfnamefont {S.}~\bibnamefont {{Weber}}}, \bibinfo
  {author} {\bibfnamefont {N.}~\bibnamefont {{Lang}}}, \bibinfo {author}
  {\bibfnamefont {H.~P.}\ \bibnamefont {{B{\"u}chler}}}, \bibinfo {author}
  {\bibfnamefont {T.}~\bibnamefont {{Lahaye}}},\ and\ \bibinfo {author}
  {\bibfnamefont {A.}~\bibnamefont {{Browaeys}}},\ }\bibfield  {title}
  {\bibinfo {title} {{Experimental realization of a symmetry protected
  topological phase of interacting bosons with Rydberg atoms}},\ }\href@noop {}
  {\bibfield  {journal} {\bibinfo  {journal} {arXiv e-prints}\ ,\ \bibinfo
  {eid} {arXiv:1810.13286}} (\bibinfo {year} {2018})},\ \Eprint
  {https://arxiv.org/abs/1810.13286} {arXiv:1810.13286 [quant-ph]} \BibitemShut
  {NoStop}%
\bibitem [{\citenamefont {Keesling}\ \emph {et~al.}(2019)\citenamefont
  {Keesling}, \citenamefont {Omran}, \citenamefont {Levine}, \citenamefont
  {Bernien}, \citenamefont {Pichler}, \citenamefont {Choi}, \citenamefont
  {Samajdar}, \citenamefont {Schwartz}, \citenamefont {Silvi}, \citenamefont
  {Sachdev}, \citenamefont {Zoller}, \citenamefont {Endres}, \citenamefont
  {Greiner}, \citenamefont {Vuleti{\'c}},\ and\ \citenamefont
  {Lukin}}]{Keesling2019}%
  \BibitemOpen
  \bibfield  {author} {\bibinfo {author} {\bibfnamefont {A.}~\bibnamefont
  {Keesling}}, \bibinfo {author} {\bibfnamefont {A.}~\bibnamefont {Omran}},
  \bibinfo {author} {\bibfnamefont {H.}~\bibnamefont {Levine}}, \bibinfo
  {author} {\bibfnamefont {H.}~\bibnamefont {Bernien}}, \bibinfo {author}
  {\bibfnamefont {H.}~\bibnamefont {Pichler}}, \bibinfo {author} {\bibfnamefont
  {S.}~\bibnamefont {Choi}}, \bibinfo {author} {\bibfnamefont {R.}~\bibnamefont
  {Samajdar}}, \bibinfo {author} {\bibfnamefont {S.}~\bibnamefont {Schwartz}},
  \bibinfo {author} {\bibfnamefont {P.}~\bibnamefont {Silvi}}, \bibinfo
  {author} {\bibfnamefont {S.}~\bibnamefont {Sachdev}}, \bibinfo {author}
  {\bibfnamefont {P.}~\bibnamefont {Zoller}}, \bibinfo {author} {\bibfnamefont
  {M.}~\bibnamefont {Endres}}, \bibinfo {author} {\bibfnamefont
  {M.}~\bibnamefont {Greiner}}, \bibinfo {author} {\bibfnamefont
  {V.}~\bibnamefont {Vuleti{\'c}}},\ and\ \bibinfo {author} {\bibfnamefont
  {M.~D.}\ \bibnamefont {Lukin}},\ }\bibfield  {title} {\bibinfo {title}
  {{Quantum Kibble--Zurek mechanism and critical dynamics on a programmable
  Rydberg simulator}},\ }\href {https://doi.org/10.1038/s41586-019-1070-1}
  {\bibfield  {journal} {\bibinfo  {journal} {Nature}\ }\textbf {\bibinfo
  {volume} {568}},\ \bibinfo {pages} {207} (\bibinfo {year}
  {2019})}\BibitemShut {NoStop}%
\bibitem [{\citenamefont {{Omran}}\ \emph {et~al.}(2019)\citenamefont
  {{Omran}}, \citenamefont {{Levine}}, \citenamefont {{Keesling}},
  \citenamefont {{Semeghini}}, \citenamefont {{Wang}}, \citenamefont {{Ebadi}},
  \citenamefont {{Bernien}}, \citenamefont {{Zibrov}}, \citenamefont
  {{Pichler}}, \citenamefont {{Choi}}, \citenamefont {{Cui}}, \citenamefont
  {{Rossignolo}}, \citenamefont {{Rembold}}, \citenamefont {{Montangero}},
  \citenamefont {{Calarco}}, \citenamefont {{Endres}}, \citenamefont
  {{Greiner}}, \citenamefont {{Vuleti{\'c}}},\ and\ \citenamefont
  {{Lukin}}}]{Omran2019}%
  \BibitemOpen
  \bibfield  {author} {\bibinfo {author} {\bibfnamefont {A.}~\bibnamefont
  {{Omran}}}, \bibinfo {author} {\bibfnamefont {H.}~\bibnamefont {{Levine}}},
  \bibinfo {author} {\bibfnamefont {A.}~\bibnamefont {{Keesling}}}, \bibinfo
  {author} {\bibfnamefont {G.}~\bibnamefont {{Semeghini}}}, \bibinfo {author}
  {\bibfnamefont {T.~T.}\ \bibnamefont {{Wang}}}, \bibinfo {author}
  {\bibfnamefont {S.}~\bibnamefont {{Ebadi}}}, \bibinfo {author} {\bibfnamefont
  {H.}~\bibnamefont {{Bernien}}}, \bibinfo {author} {\bibfnamefont {A.~S.}\
  \bibnamefont {{Zibrov}}}, \bibinfo {author} {\bibfnamefont {H.}~\bibnamefont
  {{Pichler}}}, \bibinfo {author} {\bibfnamefont {S.}~\bibnamefont {{Choi}}},
  \bibinfo {author} {\bibfnamefont {J.}~\bibnamefont {{Cui}}}, \bibinfo
  {author} {\bibfnamefont {M.}~\bibnamefont {{Rossignolo}}}, \bibinfo {author}
  {\bibfnamefont {P.}~\bibnamefont {{Rembold}}}, \bibinfo {author}
  {\bibfnamefont {S.}~\bibnamefont {{Montangero}}}, \bibinfo {author}
  {\bibfnamefont {T.}~\bibnamefont {{Calarco}}}, \bibinfo {author}
  {\bibfnamefont {M.}~\bibnamefont {{Endres}}}, \bibinfo {author}
  {\bibfnamefont {M.}~\bibnamefont {{Greiner}}}, \bibinfo {author}
  {\bibfnamefont {V.}~\bibnamefont {{Vuleti{\'c}}}},\ and\ \bibinfo {author}
  {\bibfnamefont {M.~D.}\ \bibnamefont {{Lukin}}},\ }\bibfield  {title}
  {\bibinfo {title} {{Generation and manipulation of Schr{\"o}dinger cat states
  in Rydberg atom arrays}},\ }\href@noop {} {\bibfield  {journal} {\bibinfo
  {journal} {arXiv e-prints}\ ,\ \bibinfo {eid} {arXiv:1905.05721}} (\bibinfo
  {year} {2019})},\ \Eprint {https://arxiv.org/abs/1905.05721}
  {arXiv:1905.05721 [quant-ph]} \BibitemShut {NoStop}%
\bibitem [{\citenamefont {Fisher}\ and\ \citenamefont
  {Selke}(1980)}]{Fisher1980}%
  \BibitemOpen
  \bibfield  {author} {\bibinfo {author} {\bibfnamefont {M.~E.}\ \bibnamefont
  {Fisher}}\ and\ \bibinfo {author} {\bibfnamefont {W.}~\bibnamefont {Selke}},\
  }\bibfield  {title} {\bibinfo {title} {{Infinitely Many Commensurate Phases
  in a Simple Ising Model}},\ }\href
  {https://doi.org/10.1103/PhysRevLett.44.1502} {\bibfield  {journal} {\bibinfo
   {journal} {Phys. Rev. Lett.}\ }\textbf {\bibinfo {volume} {44}},\ \bibinfo
  {pages} {1502} (\bibinfo {year} {1980})}\BibitemShut {NoStop}%
\bibitem [{\citenamefont {Bak}\ and\ \citenamefont {Bruinsma}(1982)}]{Bak1982}%
  \BibitemOpen
  \bibfield  {author} {\bibinfo {author} {\bibfnamefont {P.}~\bibnamefont
  {Bak}}\ and\ \bibinfo {author} {\bibfnamefont {R.}~\bibnamefont {Bruinsma}},\
  }\bibfield  {title} {\bibinfo {title} {{One-Dimensional Ising Model and the
  Complete Devil's Staircase}},\ }\href
  {https://doi.org/10.1103/PhysRevLett.49.249} {\bibfield  {journal} {\bibinfo
  {journal} {Phys. Rev. Lett.}\ }\textbf {\bibinfo {volume} {49}},\ \bibinfo
  {pages} {249} (\bibinfo {year} {1982})}\BibitemShut {NoStop}%
\bibitem [{\citenamefont {Bak}(1982)}]{Bak1982b}%
  \BibitemOpen
  \bibfield  {author} {\bibinfo {author} {\bibfnamefont {P.}~\bibnamefont
  {Bak}},\ }\bibfield  {title} {\bibinfo {title} {Commensurate phases,
  incommensurate phases and the devil's staircase},\ }\href
  {https://doi.org/10.1088/0034-4885/45/6/001} {\bibfield  {journal} {\bibinfo
  {journal} {Reports on Progress in Physics}\ }\textbf {\bibinfo {volume}
  {45}},\ \bibinfo {pages} {587} (\bibinfo {year} {1982})}\BibitemShut
  {NoStop}%
\bibitem [{\citenamefont {Fendley}\ \emph {et~al.}(2004)\citenamefont
  {Fendley}, \citenamefont {Sengupta},\ and\ \citenamefont
  {Sachdev}}]{Fendley2004}%
  \BibitemOpen
  \bibfield  {author} {\bibinfo {author} {\bibfnamefont {P.}~\bibnamefont
  {Fendley}}, \bibinfo {author} {\bibfnamefont {K.}~\bibnamefont {Sengupta}},\
  and\ \bibinfo {author} {\bibfnamefont {S.}~\bibnamefont {Sachdev}},\
  }\bibfield  {title} {\bibinfo {title} {Competing density-wave orders in a
  one-dimensional hard-boson model},\ }\href
  {https://doi.org/10.1103/PhysRevB.69.075106} {\bibfield  {journal} {\bibinfo
  {journal} {Phys. Rev. B}\ }\textbf {\bibinfo {volume} {69}},\ \bibinfo
  {pages} {075106} (\bibinfo {year} {2004})}\BibitemShut {NoStop}%
\bibitem [{\citenamefont {Weimer}\ and\ \citenamefont
  {B\"uchler}(2010)}]{Weimer2010}%
  \BibitemOpen
  \bibfield  {author} {\bibinfo {author} {\bibfnamefont {H.}~\bibnamefont
  {Weimer}}\ and\ \bibinfo {author} {\bibfnamefont {H.~P.}\ \bibnamefont
  {B\"uchler}},\ }\bibfield  {title} {\bibinfo {title} {{Two-Stage Melting in
  Systems of Strongly Interacting Rydberg Atoms}},\ }\href
  {https://doi.org/10.1103/PhysRevLett.105.230403} {\bibfield  {journal}
  {\bibinfo  {journal} {Phys. Rev. Lett.}\ }\textbf {\bibinfo {volume} {105}},\
  \bibinfo {pages} {230403} (\bibinfo {year} {2010})}\BibitemShut {NoStop}%
\bibitem [{\citenamefont {Sela}\ \emph {et~al.}(2011)\citenamefont {Sela},
  \citenamefont {Punk},\ and\ \citenamefont {Garst}}]{Sela2011}%
  \BibitemOpen
  \bibfield  {author} {\bibinfo {author} {\bibfnamefont {E.}~\bibnamefont
  {Sela}}, \bibinfo {author} {\bibfnamefont {M.}~\bibnamefont {Punk}},\ and\
  \bibinfo {author} {\bibfnamefont {M.}~\bibnamefont {Garst}},\ }\bibfield
  {title} {\bibinfo {title} {{Dislocation-mediated melting of one-dimensional
  Rydberg crystals}},\ }\href {https://doi.org/10.1103/PhysRevB.84.085434}
  {\bibfield  {journal} {\bibinfo  {journal} {Phys. Rev. B}\ }\textbf {\bibinfo
  {volume} {84}},\ \bibinfo {pages} {085434} (\bibinfo {year}
  {2011})}\BibitemShut {NoStop}%
\bibitem [{\citenamefont {Samajdar}\ \emph {et~al.}(2018)\citenamefont
  {Samajdar}, \citenamefont {Choi}, \citenamefont {Pichler}, \citenamefont
  {Lukin},\ and\ \citenamefont {Sachdev}}]{Samajdar2018}%
  \BibitemOpen
  \bibfield  {author} {\bibinfo {author} {\bibfnamefont {R.}~\bibnamefont
  {Samajdar}}, \bibinfo {author} {\bibfnamefont {S.}~\bibnamefont {Choi}},
  \bibinfo {author} {\bibfnamefont {H.}~\bibnamefont {Pichler}}, \bibinfo
  {author} {\bibfnamefont {M.~D.}\ \bibnamefont {Lukin}},\ and\ \bibinfo
  {author} {\bibfnamefont {S.}~\bibnamefont {Sachdev}},\ }\bibfield  {title}
  {\bibinfo {title} {{Numerical study of the chiral ${\mathbb{Z}}_{3}$ quantum
  phase transition in one spatial dimension}},\ }\href
  {https://doi.org/10.1103/PhysRevA.98.023614} {\bibfield  {journal} {\bibinfo
  {journal} {Phys. Rev. A}\ }\textbf {\bibinfo {volume} {98}},\ \bibinfo
  {pages} {023614} (\bibinfo {year} {2018})}\BibitemShut {NoStop}%
\bibitem [{\citenamefont {Whitsitt}\ \emph {et~al.}(2018)\citenamefont
  {Whitsitt}, \citenamefont {Samajdar},\ and\ \citenamefont
  {Sachdev}}]{Whitsitt2018}%
  \BibitemOpen
  \bibfield  {author} {\bibinfo {author} {\bibfnamefont {S.}~\bibnamefont
  {Whitsitt}}, \bibinfo {author} {\bibfnamefont {R.}~\bibnamefont {Samajdar}},\
  and\ \bibinfo {author} {\bibfnamefont {S.}~\bibnamefont {Sachdev}},\
  }\bibfield  {title} {\bibinfo {title} {Quantum field theory for the chiral
  clock transition in one spatial dimension},\ }\href
  {https://doi.org/10.1103/PhysRevB.98.205118} {\bibfield  {journal} {\bibinfo
  {journal} {Phys. Rev. B}\ }\textbf {\bibinfo {volume} {98}},\ \bibinfo
  {pages} {205118} (\bibinfo {year} {2018})}\BibitemShut {NoStop}%
\bibitem [{\citenamefont {Chepiga}\ and\ \citenamefont
  {Mila}(2019)}]{Chepiga2019}%
  \BibitemOpen
  \bibfield  {author} {\bibinfo {author} {\bibfnamefont {N.}~\bibnamefont
  {Chepiga}}\ and\ \bibinfo {author} {\bibfnamefont {F.}~\bibnamefont {Mila}},\
  }\bibfield  {title} {\bibinfo {title} {{Floating Phase versus Chiral
  Transition in a 1D Hard-Boson Model}},\ }\href
  {https://doi.org/10.1103/PhysRevLett.122.017205} {\bibfield  {journal}
  {\bibinfo  {journal} {Phys. Rev. Lett.}\ }\textbf {\bibinfo {volume} {122}},\
  \bibinfo {pages} {017205} (\bibinfo {year} {2019})}\BibitemShut {NoStop}%
\bibitem [{\citenamefont {Giudici}\ \emph {et~al.}(2019)\citenamefont
  {Giudici}, \citenamefont {Angelone}, \citenamefont {Magnifico}, \citenamefont
  {Zeng}, \citenamefont {Giudice}, \citenamefont {Mendes-Santos},\ and\
  \citenamefont {Dalmonte}}]{Giudici2019}%
  \BibitemOpen
  \bibfield  {author} {\bibinfo {author} {\bibfnamefont {G.}~\bibnamefont
  {Giudici}}, \bibinfo {author} {\bibfnamefont {A.}~\bibnamefont {Angelone}},
  \bibinfo {author} {\bibfnamefont {G.}~\bibnamefont {Magnifico}}, \bibinfo
  {author} {\bibfnamefont {Z.}~\bibnamefont {Zeng}}, \bibinfo {author}
  {\bibfnamefont {G.}~\bibnamefont {Giudice}}, \bibinfo {author} {\bibfnamefont
  {T.}~\bibnamefont {Mendes-Santos}},\ and\ \bibinfo {author} {\bibfnamefont
  {M.}~\bibnamefont {Dalmonte}},\ }\bibfield  {title} {\bibinfo {title}
  {{Diagnosing Potts criticality and two-stage melting in one-dimensional
  hard-core boson models}},\ }\href
  {https://doi.org/10.1103/PhysRevB.99.094434} {\bibfield  {journal} {\bibinfo
  {journal} {Phys. Rev. B}\ }\textbf {\bibinfo {volume} {99}},\ \bibinfo
  {pages} {094434} (\bibinfo {year} {2019})}\BibitemShut {NoStop}%
\bibitem [{\citenamefont {{Verresen}}\ \emph {et~al.}(2019)\citenamefont
  {{Verresen}}, \citenamefont {{Vishwanath}},\ and\ \citenamefont
  {{Pollmann}}}]{Verresen2019}%
  \BibitemOpen
  \bibfield  {author} {\bibinfo {author} {\bibfnamefont {R.}~\bibnamefont
  {{Verresen}}}, \bibinfo {author} {\bibfnamefont {A.}~\bibnamefont
  {{Vishwanath}}},\ and\ \bibinfo {author} {\bibfnamefont {F.}~\bibnamefont
  {{Pollmann}}},\ }\bibfield  {title} {\bibinfo {title} {{Stable Luttinger
  liquids and emergent $U(1)$ symmetry in constrained quantum chains}},\
  }\href@noop {} {\bibfield  {journal} {\bibinfo  {journal} {arXiv e-prints}\
  ,\ \bibinfo {eid} {arXiv:1903.09179}} (\bibinfo {year} {2019})},\ \Eprint
  {https://arxiv.org/abs/1903.09179} {arXiv:1903.09179 [cond-mat.str-el]}
  \BibitemShut {NoStop}%
\bibitem [{Note1()}]{Note1}%
  \BibitemOpen
  \bibinfo {note} {This interacting Rydberg chain is equivalent to a spin
  one-half chain with power-law antiferromagnetic Ising interactions, a
  transverse field $\protect \frac {\Omega }{2}$, and a longitudinal field
  $\Delta - \zeta (6)$. This mapping is useful to understand that the phase
  diagram features a symmetry, $(\Delta , n) \DOTSB \mapstochar \rightarrow (2
  \zeta (6) - \Delta , 1 - n)$, where $\zeta (n)$ is the Riemann Zeta
  function.}\BibitemShut {Stop}%
\bibitem [{\citenamefont {Burnell}\ \emph {et~al.}(2009)\citenamefont
  {Burnell}, \citenamefont {Parish}, \citenamefont {Cooper},\ and\
  \citenamefont {Sondhi}}]{Burnell2009}%
  \BibitemOpen
  \bibfield  {author} {\bibinfo {author} {\bibfnamefont {F.~J.}\ \bibnamefont
  {Burnell}}, \bibinfo {author} {\bibfnamefont {M.~M.}\ \bibnamefont {Parish}},
  \bibinfo {author} {\bibfnamefont {N.~R.}\ \bibnamefont {Cooper}},\ and\
  \bibinfo {author} {\bibfnamefont {S.~L.}\ \bibnamefont {Sondhi}},\ }\bibfield
   {title} {\bibinfo {title} {{Devil's staircases and supersolids in a
  one-dimensional dipolar Bose gas}},\ }\href
  {https://doi.org/10.1103/PhysRevB.80.174519} {\bibfield  {journal} {\bibinfo
  {journal} {Phys. Rev. B}\ }\textbf {\bibinfo {volume} {80}},\ \bibinfo
  {pages} {174519} (\bibinfo {year} {2009})}\BibitemShut {NoStop}%
\bibitem [{\citenamefont {Deng}\ \emph {et~al.}(2005)\citenamefont {Deng},
  \citenamefont {Porras},\ and\ \citenamefont {Cirac}}]{Deng2005}%
  \BibitemOpen
  \bibfield  {author} {\bibinfo {author} {\bibfnamefont {X.-L.}\ \bibnamefont
  {Deng}}, \bibinfo {author} {\bibfnamefont {D.}~\bibnamefont {Porras}},\ and\
  \bibinfo {author} {\bibfnamefont {J.~I.}\ \bibnamefont {Cirac}},\ }\bibfield
  {title} {\bibinfo {title} {Effective spin quantum phases in systems of
  trapped ions},\ }\href {https://doi.org/10.1103/PhysRevA.72.063407}
  {\bibfield  {journal} {\bibinfo  {journal} {Phys. Rev. A}\ }\textbf {\bibinfo
  {volume} {72}},\ \bibinfo {pages} {063407} (\bibinfo {year}
  {2005})}\BibitemShut {NoStop}%
\bibitem [{\citenamefont {Schachenmayer}\ \emph {et~al.}(2010)\citenamefont
  {Schachenmayer}, \citenamefont {Lesanovsky}, \citenamefont {Micheli},\ and\
  \citenamefont {Daley}}]{Schachenmayer2010}%
  \BibitemOpen
  \bibfield  {author} {\bibinfo {author} {\bibfnamefont {J.}~\bibnamefont
  {Schachenmayer}}, \bibinfo {author} {\bibfnamefont {I.}~\bibnamefont
  {Lesanovsky}}, \bibinfo {author} {\bibfnamefont {A.}~\bibnamefont
  {Micheli}},\ and\ \bibinfo {author} {\bibfnamefont {A.~J.}\ \bibnamefont
  {Daley}},\ }\bibfield  {title} {\bibinfo {title} {{Dynamical crystal creation
  with polar molecules or Rydberg atoms in optical lattices}},\ }\href
  {https://doi.org/10.1088/1367-2630/12/10/103044} {\bibfield  {journal}
  {\bibinfo  {journal} {New Journal of Physics}\ }\textbf {\bibinfo {volume}
  {12}},\ \bibinfo {pages} {103044} (\bibinfo {year} {2010})}\BibitemShut
  {NoStop}%
\bibitem [{\citenamefont {Nebendahl}(2015)}]{Nebendahl2015}%
  \BibitemOpen
  \bibfield  {author} {\bibinfo {author} {\bibfnamefont {V.}~\bibnamefont
  {Nebendahl}},\ }\emph {\bibinfo {title} {On the simulation of spin systems
  with tensor networks and the numerical optimization of quantum algorithms}},\
  \href {https://permalink.obvsg.at/UIB/AC11359788} {Ph.D. thesis},\ \bibinfo
  {school} {Universit\"at Innsbruck} (\bibinfo {year} {2015})\BibitemShut
  {NoStop}%
\bibitem [{\citenamefont {Pokrovsky}\ and\ \citenamefont
  {Talapov}(1979)}]{Pokrovsky1979}%
  \BibitemOpen
  \bibfield  {author} {\bibinfo {author} {\bibfnamefont {V.~L.}\ \bibnamefont
  {Pokrovsky}}\ and\ \bibinfo {author} {\bibfnamefont {A.~L.}\ \bibnamefont
  {Talapov}},\ }\bibfield  {title} {\bibinfo {title} {{Ground State, Spectrum,
  and Phase Diagram of Two-Dimensional Incommensurate Crystals}},\ }\href
  {https://doi.org/10.1103/PhysRevLett.42.65} {\bibfield  {journal} {\bibinfo
  {journal} {Phys. Rev. Lett.}\ }\textbf {\bibinfo {volume} {42}},\ \bibinfo
  {pages} {65} (\bibinfo {year} {1979})}\BibitemShut {NoStop}%
\bibitem [{\citenamefont {Haldane}(1981)}]{Haldane1981}%
  \BibitemOpen
  \bibfield  {author} {\bibinfo {author} {\bibfnamefont {F.~D.~M.}\
  \bibnamefont {Haldane}},\ }\bibfield  {title} {\bibinfo {title} {{'Luttinger
  liquid theory' of one-dimensional quantum fluids. I. Properties of the
  Luttinger model and their extension to the general 1D interacting spinless
  Fermi gas}},\ }\href {https://doi.org/10.1088/0022-3719/14/19/010} {\bibfield
   {journal} {\bibinfo  {journal} {Journal of Physics C: Solid State Physics}\
  }\textbf {\bibinfo {volume} {14}},\ \bibinfo {pages} {2585} (\bibinfo {year}
  {1981})}\BibitemShut {NoStop}%
\bibitem [{\citenamefont {Giamarchi}(2003)}]{GiamarchiBook}%
  \BibitemOpen
  \bibfield  {author} {\bibinfo {author} {\bibfnamefont {T.}~\bibnamefont
  {Giamarchi}},\ }\href {https://books.google.at/books?id=GVeuKZLGMZ0C} {\emph
  {\bibinfo {title} {Quantum Physics in One Dimension}}},\ International Series
  of Monographs on Physics\ (\bibinfo  {publisher} {Clarendon Press},\ \bibinfo
  {year} {2003})\BibitemShut {NoStop}%
\bibitem [{\citenamefont {Vidal}(2007)}]{Vidal2007}%
  \BibitemOpen
  \bibfield  {author} {\bibinfo {author} {\bibfnamefont {G.}~\bibnamefont
  {Vidal}},\ }\bibfield  {title} {\bibinfo {title} {Classical simulation of
  infinite-size quantum lattice systems in one spatial dimension},\ }\href
  {https://doi.org/10.1103/PhysRevLett.98.070201} {\bibfield  {journal}
  {\bibinfo  {journal} {Phys. Rev. Lett.}\ }\textbf {\bibinfo {volume} {98}},\
  \bibinfo {pages} {070201} (\bibinfo {year} {2007})}\BibitemShut {NoStop}%
\bibitem [{\citenamefont {{McCulloch}}(2008)}]{McCulloch2008}%
  \BibitemOpen
  \bibfield  {author} {\bibinfo {author} {\bibfnamefont {I.~P.}\ \bibnamefont
  {{McCulloch}}},\ }\bibfield  {title} {\bibinfo {title} {{Infinite size
  density matrix renormalization group, revisited}},\ }\href@noop {} {\bibfield
   {journal} {\bibinfo  {journal} {arXiv e-prints}\ ,\ \bibinfo {eid}
  {arXiv:0804.2509}} (\bibinfo {year} {2008})},\ \Eprint
  {https://arxiv.org/abs/0804.2509} {arXiv:0804.2509 [cond-mat.str-el]}
  \BibitemShut {NoStop}%
\bibitem [{\citenamefont {Kj\"all}\ \emph {et~al.}(2013)\citenamefont
  {Kj\"all}, \citenamefont {Zaletel}, \citenamefont {Mong}, \citenamefont
  {Bardarson},\ and\ \citenamefont {Pollmann}}]{Kjall2013}%
  \BibitemOpen
  \bibfield  {author} {\bibinfo {author} {\bibfnamefont {J.~A.}\ \bibnamefont
  {Kj\"all}}, \bibinfo {author} {\bibfnamefont {M.~P.}\ \bibnamefont
  {Zaletel}}, \bibinfo {author} {\bibfnamefont {R.~S.~K.}\ \bibnamefont
  {Mong}}, \bibinfo {author} {\bibfnamefont {J.~H.}\ \bibnamefont
  {Bardarson}},\ and\ \bibinfo {author} {\bibfnamefont {F.}~\bibnamefont
  {Pollmann}},\ }\bibfield  {title} {\bibinfo {title} {Phase diagram of the
  anisotropic spin-2 xxz model: Infinite-system density matrix renormalization
  group study},\ }\href {https://doi.org/10.1103/PhysRevB.87.235106} {\bibfield
   {journal} {\bibinfo  {journal} {Phys. Rev. B}\ }\textbf {\bibinfo {volume}
  {87}},\ \bibinfo {pages} {235106} (\bibinfo {year} {2013})}\BibitemShut
  {NoStop}%
\bibitem [{\citenamefont {Hauschild}\ and\ \citenamefont
  {Pollmann}(2018)}]{tenpy}%
  \BibitemOpen
  \bibfield  {author} {\bibinfo {author} {\bibfnamefont {J.}~\bibnamefont
  {Hauschild}}\ and\ \bibinfo {author} {\bibfnamefont {F.}~\bibnamefont
  {Pollmann}},\ }\bibfield  {title} {\bibinfo {title} {{Efficient numerical
  simulations with Tensor Networks: Tensor Network Python (TeNPy)}},\ }\href
  {https://doi.org/10.21468/SciPostPhysLectNotes.5} {\bibfield  {journal}
  {\bibinfo  {journal} {SciPost Phys. Lect. Notes}\ ,\ \bibinfo {pages} {5}}
  (\bibinfo {year} {2018})},\ \bibinfo {note} {code available from
  \url{https://github.com/tenpy/tenpy}},\ \Eprint
  {https://arxiv.org/abs/1805.00055} {arXiv:1805.00055} \BibitemShut {NoStop}%
\bibitem [{\citenamefont {Crosswhite}\ \emph {et~al.}(2008)\citenamefont
  {Crosswhite}, \citenamefont {Doherty},\ and\ \citenamefont
  {Vidal}}]{Crosswhite2008}%
  \BibitemOpen
  \bibfield  {author} {\bibinfo {author} {\bibfnamefont {G.~M.}\ \bibnamefont
  {Crosswhite}}, \bibinfo {author} {\bibfnamefont {A.~C.}\ \bibnamefont
  {Doherty}},\ and\ \bibinfo {author} {\bibfnamefont {G.}~\bibnamefont
  {Vidal}},\ }\bibfield  {title} {\bibinfo {title} {Applying matrix product
  operators to model systems with long-range interactions},\ }\href
  {https://doi.org/10.1103/PhysRevB.78.035116} {\bibfield  {journal} {\bibinfo
  {journal} {Phys. Rev. B}\ }\textbf {\bibinfo {volume} {78}},\ \bibinfo
  {pages} {035116} (\bibinfo {year} {2008})}\BibitemShut {NoStop}%
\bibitem [{\citenamefont {Pirvu}\ \emph {et~al.}(2010)\citenamefont {Pirvu},
  \citenamefont {Murg}, \citenamefont {Cirac},\ and\ \citenamefont
  {Verstraete}}]{Pirvu2010}%
  \BibitemOpen
  \bibfield  {author} {\bibinfo {author} {\bibfnamefont {B.}~\bibnamefont
  {Pirvu}}, \bibinfo {author} {\bibfnamefont {V.}~\bibnamefont {Murg}},
  \bibinfo {author} {\bibfnamefont {J.~I.}\ \bibnamefont {Cirac}},\ and\
  \bibinfo {author} {\bibfnamefont {F.}~\bibnamefont {Verstraete}},\ }\bibfield
   {title} {\bibinfo {title} {Matrix product operator representations},\ }\href
  {https://doi.org/10.1088/1367-2630/12/2/025012} {\bibfield  {journal}
  {\bibinfo  {journal} {New Journal of Physics}\ }\textbf {\bibinfo {volume}
  {12}},\ \bibinfo {pages} {025012} (\bibinfo {year} {2010})}\BibitemShut
  {NoStop}%
\bibitem [{\citenamefont {Nebendahl}\ and\ \citenamefont
  {D\"ur}(2013)}]{Nebendahl2013}%
  \BibitemOpen
  \bibfield  {author} {\bibinfo {author} {\bibfnamefont {V.}~\bibnamefont
  {Nebendahl}}\ and\ \bibinfo {author} {\bibfnamefont {W.}~\bibnamefont
  {D\"ur}},\ }\bibfield  {title} {\bibinfo {title} {{Improved numerical methods
  for infinite spin chains with long-range interactions}},\ }\href
  {https://doi.org/10.1103/PhysRevB.87.075413} {\bibfield  {journal} {\bibinfo
  {journal} {Phys. Rev. B}\ }\textbf {\bibinfo {volume} {87}},\ \bibinfo
  {pages} {075413} (\bibinfo {year} {2013})}\BibitemShut {NoStop}%
\bibitem [{Note2()}]{Note2}%
  \BibitemOpen
  \bibinfo {note} {We observed that particularly in simulations of crystalline
  phases, the smallest eigenvalue of the effective iDMRG-Hamiltonian does not
  coincide with the true energy of the state, although all other observables
  converge properly. Therefore, as a cross-check, we evaluate the energy with
  interactions truncated to a few hundred sites.}\BibitemShut {Stop}%
\bibitem [{\citenamefont {Pollmann}\ \emph {et~al.}(2009)\citenamefont
  {Pollmann}, \citenamefont {Mukerjee}, \citenamefont {Turner},\ and\
  \citenamefont {Moore}}]{Pollmann2009}%
  \BibitemOpen
  \bibfield  {author} {\bibinfo {author} {\bibfnamefont {F.}~\bibnamefont
  {Pollmann}}, \bibinfo {author} {\bibfnamefont {S.}~\bibnamefont {Mukerjee}},
  \bibinfo {author} {\bibfnamefont {A.~M.}\ \bibnamefont {Turner}},\ and\
  \bibinfo {author} {\bibfnamefont {J.~E.}\ \bibnamefont {Moore}},\ }\bibfield
  {title} {\bibinfo {title} {{Theory of Finite-Entanglement Scaling at
  One-Dimensional Quantum Critical Points}},\ }\href
  {https://doi.org/10.1103/PhysRevLett.102.255701} {\bibfield  {journal}
  {\bibinfo  {journal} {Phys. Rev. Lett.}\ }\textbf {\bibinfo {volume} {102}},\
  \bibinfo {pages} {255701} (\bibinfo {year} {2009})}\BibitemShut {NoStop}%
\bibitem [{Note3()}]{Note3}%
  \BibitemOpen
  \bibinfo {note} {Here we use $R=300$ in the crystalline phases and $R=10000$
  for the remaining. phases.}\BibitemShut {Stop}%
\bibitem [{\citenamefont {Fouet}\ \emph {et~al.}(2006)\citenamefont {Fouet},
  \citenamefont {Mila}, \citenamefont {Clarke}, \citenamefont {Youk},
  \citenamefont {Tchernyshyov}, \citenamefont {Fendley},\ and\ \citenamefont
  {Noack}}]{Fouet2006}%
  \BibitemOpen
  \bibfield  {author} {\bibinfo {author} {\bibfnamefont {J.-B.}\ \bibnamefont
  {Fouet}}, \bibinfo {author} {\bibfnamefont {F.}~\bibnamefont {Mila}},
  \bibinfo {author} {\bibfnamefont {D.}~\bibnamefont {Clarke}}, \bibinfo
  {author} {\bibfnamefont {H.}~\bibnamefont {Youk}}, \bibinfo {author}
  {\bibfnamefont {O.}~\bibnamefont {Tchernyshyov}}, \bibinfo {author}
  {\bibfnamefont {P.}~\bibnamefont {Fendley}},\ and\ \bibinfo {author}
  {\bibfnamefont {R.~M.}\ \bibnamefont {Noack}},\ }\bibfield  {title} {\bibinfo
  {title} {Condensation of magnons and spinons in a frustrated ladder},\ }\href
  {https://doi.org/10.1103/PhysRevB.73.214405} {\bibfield  {journal} {\bibinfo
  {journal} {Phys. Rev. B}\ }\textbf {\bibinfo {volume} {73}},\ \bibinfo
  {pages} {214405} (\bibinfo {year} {2006})}\BibitemShut {NoStop}%
\end{thebibliography}
%

\end{document}